\documentclass[]{article}
\usepackage{emulateapj}
\usepackage{graphicx}
\usepackage[usenames,dvips]{color}

\advance \voffset by -0.5cm\relax

\def\msun{{\rm ~M}_{\odot}}

\def\mpy{{\rm ~M}_{\odot} {\rm ~yr}^{-1}}

\begin{document}

\title{Short-Hard Gamma-Ray Bursts in Young Host Galaxies: the Effect of Prompt
Twins}

\author{Krzysztof Belczynski\altaffilmark{1,2,3}, 
         K. Z. Stanek\altaffilmark{4},
         Chris L. Fryer\altaffilmark{1,5}}

\affil{
     $^{1}$ Los Alamos National Lab, 
            P.O. Box 1663, MS 466, Los Alamos, NM 87545 \\
     $^{2}$ Oppenheimer Fellow \\
     $^{3}$ New Mexico State University, Dept of Astronomy,
            1320 Frenger Mall, Las Cruces, NM 88003\\
     $^{4}$ Department of Astronomy, Ohio State University, Columbus, OH 43210\\
     $^{5}$ Physics Department, University of Arizona, Tucson, AZ 85721\\
     kbelczyn@lanl.gov, kstanek@astronomy.ohio-state.edu, clfreyer@lanl.gov}

\begin{abstract}

We investigate the effect of including a significant ``binary twin''
population (binaries with almost equal mass stars, $q = M_{\rm 2}/M_{\rm 1} 
\gtrsim 0.95$) for the production of double compact
objects and some resulting consequences, including LIGO inspiral rate
and some properties of short-hard gamma-ray bursts.  We employ very
optimistic assumptions on the twin fraction ($\sim50\%$) among all
binaries, and therefore our calculations place an upper limits on the
influence of twins on double compact object populations. We show that
for LIGO the effect of including twins is relatively minor: although the 
merger rates does indeed increase when twins are considered, the rate
increase is fairly small ($\sim1.5$). Also, chirp mass distribution
for double compact objects formed with or without twins are almost
indistinguishable. If double compact object are short-hard GRB
progenitors, including twins in population synthesis calculations does
not alter significantly the earlier rate predictions for the event
rate. However, for one channel of binary evolution, introducing twins
more than doubles the rate of ``very prompt'' NS-NS mergers (time to
merger less than $10^{6}$ years) compared to models with the ``flat''
$q$ distribution. In that case, $70\%$ of all NS-NS binaries merge
within $10^8$ years after their formation, indicating a possibility of
a very significant population of ``prompt'' short-hard gamma-ray
bursts, associated with star forming galaxies.  We also point out
that, independent of assumptions, fraction of such prompt
neutron star mergers is always high, $\sim35-70\%$.  We note that
recent observations (e.g., Berger et al.)  indicate that fraction of
short-hard GRBs found in young hosts is at least $\sim40\%$ and
possibly even $\sim80\%$.

\end{abstract}


\section{Introduction}

A majority of stars are in binaries, and a substantial fraction of
binaries have short enough orbital periods that they are likely to
interact during either their main sequence or post-main sequence
evolution.  Many of the most interesting phenomena in astronomy can be
directly traced to the interaction of close binaries; an incomplete
list would include binary neutron stars and white dwarfs, supernovae
Ia, cataclysmic variables, and blue stragglers.  There is a vast
literature on the subject (e.g., Paczynski 1971; Wellstein \& Langer
1999; Hurley, Tout \& Pols 2002; Belczynski, Kalogera \& Bulik 2002b).
Although there are many ingredients that must be considered in
interacting binaries, an implicit assumption in much theoretical work
has been that the lifetimes of the stars are almost always quite
different. This assumption arises naturally from two considerations.
First, the single star initial mass function (IMF) is a steep function 
of mass, with low mass
stars being far more numerous than high mass stars (e.g. Salpeter
1955), and strong mass-lifetime relationship for all but the most
massive stars implies a large lifetime difference unless the masses
are very close.  Second, a flat mass ratio spectrum (see for example
Kuiper 1935) for binaries that are likely to interact is adopted in
most population synthesis studies, leading to very few ``equal''
component mass binaries.

Pinsonneault \& Stanek (2006) have argued that observations indicate
the existence of a substantial population of nearly equal mass
binaries (``twins'').  In such systems a strong inequality in lifetime
is not present, so there might be important qualitative differences in
their evolution compared to unequal mass binaries. Survey of
astronomical literature strongly suggests binary twins are a general
feature of close binary population, as a peak near $q=1$ was reported
by a number of investigators. For example, Halbwachs et al. (2003)
studied a large sample of spectroscopic binaries type F7 to K (masses
from about 1.7 down to $0.5 \msun$), including binaries in open
clusters.  They find that the mass ratio has a broad peak from $q
\approx 0.2$ to $q \approx 0.7$, and a sharp peak for $q>0.8$. As they
discuss, the strength of the peak for $q>0.8$ gradually decreases with
the increasing orbital period, which is to be expected.  The fraction
of twins can be as high as $50\%$ for periods $P<10\;$days and it is
still significant (as high as 35\%) for much longer periods of up to
1000 days. A much earlier study by Lucy \& Ricco (1979) also finds a
strong and narrow peak of binaries with $q \approx 0.97$, again using
a sample of spectroscopic binaries corrected for various observational
errors and biases. Tokovinin (2000) confirms that finding using
additional data and in fact also calls this population ``twins'',
arguing that they constitute 10-20\% of the total binary population in
the $P=2-30$ days regime.

Additional, although perhaps more anecdotal support for the
significant twin population comes from the realm of very high mass
stars found in eclipsing binaries. The most massive binary known, WR
20a (Rauw et al. 2004; Bonanos et al.  2004), is an eclipsing system,
so the masses of both components can be measured accurately.  The
masses are $83\;M_{\odot}$ and $82\;M_{\odot}$ (Rauw et al. 2005),
giving a mass ratio of $q=0.99$.  Given that $80\;M_{\odot}$ stars are
extremely rare (both due to the steepness of the mass function and
their short lifetime), having such extremely massive secondary would
be quite unlikely unless the twin phenomenon is involved.

There are also some theoretical considerations that seem to indicate
that double neutron star binaries form {\em only} from twins 
(Bethe \& Brown 1998; Chang-Hwan, Hong-Jo \& Brown 2007).  If this is 
the only double neutron star formation scenario, the twin fraction must 
be high to explain the observed rates of these binary systems.  

However, not all evidence points towards a large population of twins.
First, there are some loopholes to the arguments pushing toward the
theoretical requirement of twins to make double neutron star systems.
In addition, the existence of low-mass X-ray binaries requires some
systems with very different masses (Kalogera \& Webbink 1998; Fryer, 
Burrows \& Benz 1998). Even with the intermediate-mass progenitors 
of these low-mass X-ray binaries (Podsiadlowski, Rappaport \& Pfahl 
2002), a large twin fraction coupled on top of a otherwise flat mass 
ratio distribution would have trouble explaining low-mass X-ray binaries.  
Finally, not all the observational evidence points toward a twin fraction.
Kobulnicky \& Fryer (2007) argue that for their dataset of 120 O and
early B stars, the twin fraction must be less than 25\%.  Their study
used one of the largest datasets of O and early B stars focusing on 
a single stellar association - Cygnus OB2 (Kiminki et al. 2007).  

With observations and theory arguing both for and against twins, we
investigate the effect of twin binaries on population of close
(coalescing within Hubble time) double compact objects, focusing on
observations that might allow us to distinguish a twin population of
stars from the more standard stellar mass ratio distributions. In this 
study we present the population synthesis study of double neutron star 
(NS-NS), black hole neutron star (BH-NS) and double black hole (BH-BH) 
progenitors.  We employ
two basic calculations; one with the usually adopted flat mass ratio
distribution and one that includes a very large ($50\%$) population of
twins. The results are discussed in context of double compact object
mergers that are expected to be the prime sources of gravitational
radiation for ground based observatories like LIGO or VIRGO (e.g.,
Kalogera et al. 2007), and are also considered as very likely
short-hard gamma ray burst progenitors (Nakar 2007). In a forthcoming
paper (Belczynski \& Pinsonneault, in prep.)  we will study the
influence of twins on lighter compact object binaries with white
dwarfs and their connection to Type Ia supernovae and formation of
blue stragglers.

\section{Model}

\subsection{Population synthesis model}  

Binary population synthesis is used to calculate the merger rates 
and properties of double compact objects. The population synthesis 
code employed in this work, {\tt StarTrack}, was initially developed 
for the study of double compact object mergers in the context of 
gamma-ray burst (GRB) progenitors (Belczynski, Bulik \& Rudak 2002a) 
and gravitational-wave inspiral sources (Belczynski et al. 2002b). 
In recent years {\tt StarTrack} has undergone major updates
and revisions in the physical treatment of various binary evolution
phases, and especially mass transfer phases. The new version has
already been tested and calibrated against observations and detailed
binary mass transfer calculations (Belczynski et al.\ 2007a), and has
been used in various applications (e.g., Belczynski, Bulik \& Ruiter
2005; Belczynski et al. 2006; Belczynski et al.\ 2007b). The physics
updates that are most important for compact object formation and
evolution include: a full numerical approach to orbital evolution due
to tidal interactions, calibrated using high mass X-ray binaries and
open cluster observations, a detailed treatment of mass transfer
episodes fully calibrated against detailed calculations with a stellar
evolution code, updated stellar winds for massive stars, and the
latest determination of the natal kick velocity distribution for
neutron stars (Hobbs et al.\ 2005). For helium star evolution, which
is of a crucial importance for the formation of double neutron star
binaries (e.g., Ivanova et al.\ 2003; Dewi \& Pols 2003), we have
applied a treatment matching closely the results of detailed
evolutionary calculations. If the helium star fills its Roche lobe,
the systems are examined for the potential development of a dynamical
instability, in which case they are evolved through a common envelope 
(CE) phase,
otherwise a highly non-conservative mass transfer issues. We treat CE
events using the energy formalism (Webbink 1984), where the binding
energy of the envelope is determined from the set of helium star models
calculated with the detailed evolutionary code by Ivanova et al.\
(2003). The progenitor evolution and the Roche lobe overflow episodes
are now followed in much greater detail. We note significant
differences from our earlier studies.  For a detailed description of
the revised code we refer the reader to Belczynski et al.\ (2007a).

\subsection{Recent model revisions}  

The most recent and important changes
in the context of double compact object formation reflect the
treatment of the dynamically unstable mass transfer and evolution into
the CE phase. First, it was pointed out that there is only (if any) a
small chance of survival of CE phase if a donor star is on the
Hertzsprung gap (HG), simply because there is no clear entropy jump
between core and envelope so once CE inspiral is initiated it does not
stop until the two binary components coalesce (see Belczynski et al.
2007b).  Second, we limit accretion onto compact objects during CE
phase to $10\%$ of the Bondi-Hoyle rates based loosely on estimates of
outflows (Armitage \& Livio 2001).  We have also slightly modified our
input physics in context of rejuvenation, black hole spin (Belczynski
et al. 2007c) evolution and debugged the entire code.

\subsection{Calculations}  

Two separate evolutionary models for massive star binaries are 
calculated. They differ only in common envelope treatment. In one 
calculation (model A) that we will refer to as our reference model 
we do not allow for common envelope survival in case the donor star is 
crossing HG. This is in effect for H-rich HG stars as well of helium 
HG stars. The former reduces drastically formation (and merger) rates 
of BH-BH binaries, while the later reduces moderately rates for NS-NS 
systems as discussed in detail by Belczynski et al (2007b). In 
alternative common envelope model (model B) we allow for CE survival for 
all donors (HG included). It does not mean that system can survive every
CE phase. The regular standard energy balance (e.g. Webbink 1984) is
performed to check for a potential survival.  In both models we vary
an assumption on the initial mass ratio (lower-mass over higher-mass
binary component) of binaries that we evolve. We either employ flat
mass ratio distribution and we will refer to these populations as
``flat'' binaries or we impose ``twin'' distribution in which we
require that $50\%$ binaries have mass ratio distributed uniformly in
range $q=0.95-1.0$ while the remaining $50\%$ have flat distribution for 
$q=0.0-0.95$. For each models we evolve $N_{\rm tot} = 1.465
\times 10^6$ binaries with solar metallicity ($Z=0.02$).  We require
that the primary mass is drawn from power-law IMF with slope $-2.7$,
while secondary mass is obtained through a given mass ratio
distribution.  We additionally require that the primary initial mass
is $6>M_1>150 \msun$ while secondary initial mass is $4>M_2>150
\msun$. The range of masses was chosen such that it encompasses entire
possible mass range for double compact object formation. In
particular, low-mass ends take into account potential rejuvenation of
stars through binary accretion. To initialize our populations we first
draw a primary mass, then mass ratio is drawn from a given
distribution, and then mass of a secondary is calculated from $M_2=q
\times M_1$. If $M_2$ is smaller than required minimum mass ($M_2 = 4
\msun$) we repeat the drawing. Such a scheme, although it uses
underlying flat mass ratio distribution results in skewed (toward
high-$q$) distribution. The resulting initial mass ratio distributions
are presented in Figure 1 (top panel).

\subsection{Calibration} 

For calibration and Galactic compact object merger
rate calculation we use binary fraction of $f_{\rm bi}=0.5$, and we
assume that star formation rate (SFR) was constant in Galaxy through
last 10 Gyr at the level of $3.5 \mpy$.  To calculate the synthetic
SFR we extend our IMF down to hydrogen burning limit ($0.08 \msun$),
with a three component broken power-law IMF with slopes of
$-1.3/-2.2/-2.7$ and corresponding breaks at $0.5 \msun$ and $1.0
\msun$ (Kroupa \& Weidner 2003). For our twin populations we assume
that twin binaries are formed independent of binary properties (like
period) and that they form in entire mass range ($0.08>M>150
\msun$). The mass of entire underlying stellar population (all single
and binary stars) that corresponds to our simulations ($F_{\rm sim}$)
is $7.729 \times 10^8 \msun$ and $6.182 \times 10^8 \msun$ for flat
and twin populations, respectively. Since the star forming mass in
Galaxy is $F_{\rm sfr} = 3.5 \times 10^{10} \msun$ it results in
calibration boost factors ($F_{\rm x}=F_{\rm sfr}/F_{\rm sim}$) of
$F_{\rm x}=45$ and 57 for flat and twin populations,
respectively. After evolution of massive primordial binaries ($N_{\rm
tot}=1.465 \times 10^6$) we obtain population of double compact
objects in each model. Then in a given model we initiate each double
compact object $F_{\rm x}$ times at different starting time. Starting
times are chosen from the uniform distribution within the range of
$0-10$ Gyr (constant SFR). The starting time is then increased by an
evolutionary time that was needed for a progenitor binary to form a
given double compact object ($T_{\rm evol} \sim10-20$ Myr). The
double compact objects are then evolved with angular momentum losses
due to emission of gravitational radiation until they merge. Merger
times are denoted as $T_{\rm mer}$ and they can span a wide range of
values. The entire lifetime of a given binary is then $T_{\rm
life}=T_{\rm evol}+T_{\rm mer}$. We record the time at which they
merge. Then we calculate an average merger rate in period $0-10$
Gyr. These are our predicted Galactic merger rates. It is worth to
note three things.

First, it may seem counter-intuitive that the boost factor is larger
for (more massive) twin population. However, one needs to realize that
in the population of stars of a given mass there is a larger number of
high mass binaries ($6>M_1>150 \msun$ and $4>M_2>150 \msun$) in twin
population as compared with flat population. Simply, it is easier to
form both binary components with high masses in a population with mass
ratio peaked at high values (twins) as opposed to population with
flatter mass ratio distribution.  As we have evolved the same number
of twins and flat high mass binaries, it means that the number of
stars in an entire underlying stellar population ($0.08-150 \msun$) is
smaller for twins than for flat binaries. Finally, since the most mass
is contained in primaries and single stars, and not in secondaries
(that are heavier in twin population), it translates into a smaller
mass of underlying stellar population containing twin binaries.
Smaller the simulated mass ($F_{\rm sim}$) higher the boost factor.

Second, we have employed an optimistic (pro-twin) approach, as we do
not put any period constraints on twin formation (see \S\,1 discussing
evidence that twins may form preferentially at short periods) in
addition to adopting a very high fraction of twins ($50\%$). Had we
limited population of twins, the boost factor $F_{\rm x}$ for twins
would decrease, making the differences between twin and flat
calculations less pronounced.

Third, since, we also consider population of ultracompact (extremely
short-lived) double neutron star binaries it is important to notice
their increased contribution to merger rates. If at formation there
are similar in size populations of short- and long-lived double
compact objects, the short-lived systems will dominate merger
rates. Long-lived systems merge beyond our counting time of 10 Gyr (age
of the disk) unless they happen to form early on, while short-lived
systems contribute to merger rate independent of their formation time
(provided that their merger times are much shorter than the age of the
disk).

\section{Results}

\subsection{Rates} 

First we have calculated Galactic merger rates for the two
models and we have translated them into advanced LIGO detection rates
using method presented in Belczynski et al. (2007b). The results are
presented in Table~1.

The Galactic merger rates of double compact objects (combined for
NS-NS/BH-NS/BH-BH) for flat populations are $13-48$ Myr$^{-1}$ while
for twin populations $20-74$ Myr$^{-1}$. The range of the rates
corresponds to our different assumption on CE evolution and formation
(or lack of thereof) of ultracompact NS-NS systems as was discussed in
detail by Belczynski et al. (2007b). The factor of $\sim1.5$
increase in rates from flat to twin dominated populations is equally
connected to {\em (i)} the difference in underlying star population
that gives boost factor $\sim25\%$ larger for twins than for flat
binaries (see SFR calibration \S\,2.4) and {\em (ii)} the slightly
higher ($\sim20\%$) formation efficiency of double neutron stars from
massive twin binaries. The small magnitude of the later effect may be
somewhat surprising, as one would intuitively expect that with the
twin population production of double compact objects would
significantly increase.  In the following we explain this surprising
finding.
 
First, we examine the mass ratio distribution of flat population for
model A.  There is a significant fraction ($\sim40\%$) of massive
binaries that we have evolved with low mass ratios ($q_{\rm init} <
0.65$; Fig.1; top panel).  On the other hand, binaries that actually
produce double compact objects (Fig.1; middle panel) are found
predominantly with high mass ratios but in a rather wide range
($q_{\rm init} \sim0.65-1$). Note that there is no significant peak
for double compact object progenitors at high-$q_{\rm init}$.  Second,
if we go from flat to twin population we shift half of the initial
binaries from the entire mass ratio range to the very high mass ratios
(Fig. 1; top panel). Binaries that are shifted from the low-$q$ range
($\sim20\%$ of the population) will become an extra component in
formation of double compact object in twin population. Binaries that
are shifted from the high-$q$ range will produce double compact
objects but will not increase the overall production rate since they
would have formed double compact objects anyway. Therefore, the rate
increase factor from the shift of binaries from standard to twin
population is only $\sim20\%$.

The above finding is a direct result of the shape of the mass ratio
distribution of double compact object progenitors. In model A for the
flat population mass ratio is found within range $q_{\rm init} \sim
0.5-1$ and it falls slowly with the decreasing $q_{\rm init}$. The
lack of progenitors below $q_{\rm init} \sim0.5$ is connected to fact
that below that value the progenitor binary evolves through common
envelope (rather than stable mass transfer phase) after primary
evolved of main sequence, and the CE leads most often to a
merger. This is especially true since most of the donors will start
mass transfer on Hertzsprung gap as during this phase stars experience
maximum radial expansion (Belczynski et al.2007b).\footnote{We can see
that if we relax our assumption on CE mergers in model B (Fig.1,
bottom panel) the mass ratio of double compact progenitors extends to
low-$q_{\rm init}$ values.}  The slope of the distribution is
explained by the narrow range of masses in which double neutron stars
form (and since they dominate double compact object population they
set the distribution). If a primary is chosen within a range for NS
formation ($\sim8-20 \msun$) it is easier to find potential secondary
that can form NS if mass ratio is higher. If mass ratio is too small,
the primary have a greater chance to have mass below NS formation
mass, and therefore mass ratio distributions falls off with decreasing
$q_{\rm init}$. Intrinsically, once two stars have masses within NS
formation range and their mass ratio is over 0.5, there is no
preference for NS-NS formation at higher $q$. In other words, we do
not note any significant evolutionary effects that make it easier to
make NS-NS at high mass ratio.

Advanced LIGO detection rates are listed in Table 1. As for Galactic
merger rates there is a range of values for flat population: $15-700$
yr$^{-1}$ and for twin population $22-825$ yr$^{-1}$. And as before
the range corresponds to the change on assumption on CE
evolution. However, the increase in rates from model A to B is now due
to the increased formation of BH-BH binaries in model B. These
binaries, although a small contributor to Galactic merger rates, are
most important for LIGO as they can be detected from much larger
distances (much higher chirp masses) as compared to NS-NS mergers and
therefore they dominate detection rates (see also Belczynski et
al. 2007b).  We note that the change of the detection rates for LIGO
from flat to twin population is rather small (factor of $\sim1.5$)
and is much smaller than other model uncertainties (e.g. CE evolution).

\subsection{Double compact object chirp mass}  

In Figure 2 we show the
distribution of chirp mass for coalescing double compact objects. As
the population of double compact objects is dominated by double
neutron stars we see that the distributions peak at $\sim 1.2 \msun$
that corresponds to the typical chirp mass of a $1.35$ and $1.35
\msun$ NS-NS binary (see also Belczynski et al. 2007d). We also note
that the distributions are almost the same for the flat and twin
populations. This is the result of the underlying initial final-mass
relation (see Belczynski et al. 2007a for details).  This relation
shows that neutron stars form with the similar mass ($\sim 1.35
\msun$) for a wide range of progenitor masses $M_{\rm zams} \sim 8-18
\msun$ and only in the narrow range $M_{\rm zams} \sim 18-20 \msun$
heavier neutron stars ($\sim 1.8 \msun$) are formed.  Such the
initial-final mass relation leads to a rather narrow distribution of
neutron star masses (somewhat widened by accretion and mass loss in
binaries) that is obtained in both populations. If the two populations
are compared in context of the flat initial-final mass relation it
becomes obvious why the two distributions peak at the same value. For
flat population two neutron star progenitors are found (on average)
farther apart in mass than for twins but still they need to fall
within the narrow mass limits that allow neutron star formation
($M_{\rm zams} \sim 18-20 \msun$). For twin model the two progenitors
are closer in mass, but still are within the same mass limits. Since
the mass of a neutron star does not depend strongly on the initial
mass of progenitor the masses of neutron stars in both models are
similar.  There are other heavier compact objects, namely black hole
neutron star systems and double black hole systems with chirp masses
reaching all the way to $\sim 10 \msun$, both for twin and flat
populations. In particular we find many more heavier systems in model
B as in this model black hole systems form with much higher efficiency
as compared to model A (Belczynski et al. 2007b).

\subsection{Merger times}  

Merger time distributions for flat and twin
double compact object binaries are presented in Figure 3. In the top
panel we show calculations with our reference evolutionary model, while
the bottom panel demonstrates results for the alternative common
envelope evolution.  For the reference model the two distributions are
very similar, and the number of mergers falls off rapidly with the
decreasing merger time.  However, we still predict quite a significant
number of double compact objects: $\sim 35\%$ with merger times
shorter than 100 Myr.  Most of these short lived systems are double
neutron stars that have formed along evolutionary channels that end 
in the stable mass transfer episode with a helium star donor
(e.g., Ivanova et al. 2003). In the model with alternative
evolution we allow for common envelope survival even if donor stars
are crossing Hertzsprung gap. Although this may appear not to be
supported by the current understanding of inspiral process (see \S\,2
and Belczynski et al. 2007b for more through discussion) the common
envelope evolution and the associated inspiral is not yet fully
understood.  Distributions are similar for flat and twin binaries for
high merger times. However, for small merger times there is a an
additional component in both distributions as compared to the standard
calculations.  Moreover, this additional component is much more
pronounced in twin population than in flat population. In particular
we find that in flat population this component ($T_{\rm mer} \lesssim 
1$ Myr) contains $33\%$ of mergers while in twin population it reaches
$50\%$. Accounting for the shape of distribution and the larger number
of mergers in twin population it translates to $\sim 2.5$ times as
many short-lived systems in twin population as compared to flat
population.  The systems with very short merer times ($T_{\rm mer} 
\lesssim 1$ Myr) are so called "ultracompact" double neutron stars,
that form through one extra common envelope phase (additional orbit
contraction) as opposed to standard model binaries with larger merger
times (e.g., Belczynski et al. 2002b; Ivanova et al. 2003; Belczynski
et al. 2006).

In Table~1 we list fractions of prompt double compact object mergers.
These will include potential short-hard GRB progenitors: NS-NS and
BH-NS mergers. Although, it is noted again that these fractions are
almost completely dominated by NS-NS mergers. Fractions are given for
binaries that have lifetimes ($T_{\rm evol}+T_{\rm mer}$) shorter than
100 ($F_{\rm 100}$) and 1000 Myr ($F_{\rm 1000}$). We find that in the
reference model $\sim 35\%$ of the mergers are expected to occur in
young hosts (with stellar populations as young as 100 Myr) both for
flat and twin models. However, if alternative evolution is included in
calculations, then the fraction increases to $60\%$ for flat
population and to $70\%$ for twin population. This is a direct result
of merger time distribution that is similar for twin and flat
population in the reference model (see Fig.~3 top panel) and different
for alternative CE model, in particular twins producing many more
ultracompact NS-NS binaries (see Fig.~3 bottom panel).
 
The fractions of the mergers are also given in Table 1 for
significantly older (but still rather young) hosts: 1000 Myr. It is
found that great majority of the mergers $\sim 70\%$ and $\sim 80\%$
for reference and alternative CE models is then expected to take
place in hosts of this age. At this age (or lifetime of the double
compact object population) the ultracompacts are not so important as
the classical long-lived systems play an important role in overall
population and the fractions are rather independent of whether twin of
flat populations are considered (see Fig.~3).

In the following we explain the more effective production of
ultracompacts in the twin population than in the flat population for
alternative CE evolution (see Fig.~3; model B).  We will limit 
the discussion to double neutron stars and their progenitors as they 
constitute the vast majority ($\sim 99\%$) of double compact object 
(DCO) systems with ultrashort merger times (i.e., $T_{\rm mer}<1$ Myr). 
The distribution of initial mass ratio for progenitors of ultracompact 
DCOs is presented in Figure~4.

For the flat population (see Fig.~4, top panel) we notice that {\em
(i)} in model A there are rather few progenitors with high mass ratios
($q_{\rm init} =0.95-1$) in contrast to model B in which we find a
prominent peak of the distribution at high mass ratios. Therefore for
model B, redistribution of progenitors from the flat to twin mass
ratio distribution is enhancing the production of ultracompacts. In
fact, for twin population (see Fig.~4, bottom panel) we observe an
increase of ultracompact systems by a factor of $\sim2.5$ in model B
(see the significant increase of these systems at high-$q_{\rm
init}$). Note that the change of the mass ratio distribution from flat
to twin increases number of progenitors due to the calibration (see
\S\,2) but this increase factor is only $\sim1.2$. The additional
increase is solely due to the peak in the number of high-$q_{\rm
init}$ systems for model B ultracompacts.

The shape of the mass-ratio distribution for progenitors of
ultracompact systems is understood in the framework of evolution of
massive stars leading to the formation of double neutron stars. In
general, classical (long-lived) NS-NS binaries form from progenitors
that experience only two mass transfer episodes. The ultracompact
systems progenitors usually go through an additional mass transfer
episode. This third mass transfer episode is encountered just before
second NS formation. A low-mass helium star overfills its Roche lobe
and starts transferring He-rich material to the first born NS. Most
often such a transfer occurs when a helium star is crossing
Hertzsprung gap (large radial expansion). Depending on the mass ratio
and the evolutionary stage of the helium donor (where on HG) the mass
transfer is either stable or it evolves into CE phase. Since the most
of the neutron stars in our simulations have mass $1.3-1.4 \msun$ the
mass ratio is set by the mass of helium donor. For very light helium
stars ($\sim3 \msun$), stable mass transfer is predicted while, for
more massive donors ($\gtrsim 3.5 \msun$), a CE develops. The mass of
the helium star is set predominantly by the initial mass of the
progenitor star (in addition to mass gain and loss in earlier binary
interactions); the lower the mass of the progenitor, the lower mass of
helium star it forms. Since the helium star is formed out of secondary
(most cases) and $q_{\rm init}$ was defined as the ratio of the mass
of secondary to primary, it is expected that systems going through the
stable mass transfer (lower mass helium star progenitor) have
initially a lower mass ratio than systems developing CEs (higher mass
helium star progenitor).

In model A we do not allow for CE survival if donor is on HG and
therefore progenitors with very high mass ratios are disfavored.  In
model B, that allows for survival of the CE phase with the HG donor,
the high mass ratio progenitors are abundant and they contribute to
formation of ultracompacts. Additionally, the higher mass ratio
systems are more likely to survive initial mass transfer episodes in
the evolution of progenitor binary (e.g., closer in mass components so
lower the probability of a merger during CE phase).

\section{Comparison with earlier studies}

Bethe \& Brown (1998) proposed a scenario of double neutron star
formation from twin binaries.  In this scenario, because the two
stellar components of the binary are nearly equal mass, the system
undergoes both common envelope phases prior to the collapse of either
star.  In such a scenario, the neutron stars formed in collapse need
not undergo a CE phase, and hence avoid accreting additional material.
This model provides a natural explanation for double neutron star
systems in which both neutron stars had nearly equal masses.  But it
only works when the two binary components have nearly equal mass, 
and hence, strongly depends on the number of twins.  Bethe \& Brown (1998)
and subsequent work by Lee et al. (2007) argue that this scenario can
explain all of the NS-NS binary systems observed if a large twin 
population exists.  They argue that any formation
scenario that forces a neutron star to go through a common envelope
phase will produce a low-mass black hole, not a neutron star.  The
bulk of the simulations by Fryer, Woosley \& Hartmann  (1999) also made this
assumption, and came up with similar conclusions:  with appropriate
choices of the other free parameters, one can match the observed NS-NS
systems.

But whether or not this scenario is the dominant formation path for
double neutron star binaries hinges on the fact that the accretion
onto a neutron star in a common envelope system is equal to the
Bondi-Hoyle rate.  Recall that it was realized that neutrino cooling
would allow the neutron star to accrete beyond the Eddington rate,
causing the neutron star to accrete as much material as is fed it.  If
one assumes this rate is equal to the Bondi-Hoyle rate, the accretion
can be up to a solar mass.  But the actual accretion rate may be much
less.  First, how much one accretes depends sensitively on the evolution 
of the common envelope phase (Fryer et al. 1999). Very few
simulations have focused particularly on neutron stars in stellar
mergers with giant companions.  Most examples have very rough boundary
conditions and/or do not model the inspiral of a neutron star in a
massive companion (e.g. Ruffert 1999; Armitage \& Livio 2000; Zhang \&
Fryer 2001; Ricker \& Taam 2007).  This preliminary work has yet to
solve the actual merger process.  But these studies have determined a
few key issues with the Bondi-Hoyle assumption in neutron star
accretion in stellar inspiral: density/velocity gradients can alter
the Bondi-Hoyle accretion at large scales, density/velocity gradients
can lead to disk formation and outflows.

Ruffert \& Anzer (1995), Ruffert (1999) and Taam \& Ricker (2007) have
focused on the deviation at large scales of the Bondi-Hoyle accretion
rate.  At issue here is that angular momentum in the accreting
material can provide pressure support for the infalling material,
slowing the accretion.  Fryer et al. (1996) calculated values 
for the density and velocity gradients and compared these results 
with those of Ruffert \& Anzer (1995) and found that this pressure 
support on the global scale was minimal ($< 40$\% level for quite 
large velocity gradients).  Since this time, Ruffert (1999) studied 
the same effect under density gradients.  Again, if we use the 
estimates from Fryer et al. (1996) for the density gradients in 
10,20\,M$_\odot$ supergiants, we expect 10-20\% variations away 
from the accretion rate predicted by the Bondi-Hoyle formalism.  
This only changes as the neutron star spirals near the inner 
edge of the hydrogen envelope, where density gradients can become 
quite large.

Fryer, Benz \& Herant (1996) assumed that as long as Bondi-Hoyle accretion 
were unaffected, the angular momentum would somehow be transported outward,
allowing the material to accrete onto the neutron star.  But
subsequent studies are showing that this assumption may well be
incorrect (Armitage \& Livio 2001; Fryer et al. 2006; Fryer 2007).
Armitage \& Livio (2001) showed that the angular momentum in the
inflow would lead to disk formation, and ultimately, an outflow that
could halt accretion.  If the material is unable to get rid of the
energy produced by viscous interactions, an outflow is bound to occur
(Blandford \& Begelman 1999).  Fryer et al. (2006) and Fryer (2007)
have specifically studied accreting neutron star systems and found
that even a small amount of angular momentum would lead to outflows.
In these low-angular momentum flows, the outflows decreased the
accretion rate by 50\%, but for the high-angular momentum flows in CE
phases, the outflow could decrease the accretion by more than an order
of magnitude (Blandford \& Begelman 1999).  Because of such results,
we estimate our mass accretion by assuming an accretion rate of 10\%
the Bondi-Hoyle rate.  It could even be an order-of-magnitude lower.

This reduced accretion rate allows additional scenarios for forming
double neutron star systems which can also be shown to match the
current observations of double neutron star systems (this study;
Belczynski et al. 2007d).  It also avoids any problems over-producing
(or hiding) the number of massive NS-NS systems.  Brown
\& Bethe (1998) turned these systems into low-mass BH-NS systems by
requiring a maximum neutron star mass between 1.7-1.8\,M$_\odot$.
However, observations may indicate that the maximum neutron star mass
is $M_{\rm ns,max} \sim 2 \msun$ (e.g., Ransom et al. 2005; Barziv et
al. 2001) while some theoretical work allows for equation of states
with $M_{\rm ns,max} \sim2-3 \msun$ (e.g., Morrison, Baumgarte \&
Shapiro 2004).  If these systems form black holes, we also do not see
the low-mass black holes in the Galaxy - all black holes in the Galaxy
have masses above $\sim3 \msun$ (Orosz 2003; Cesares 2007), although 
Fryer \& Kalogera (2001) argued that this is more-likely the result of
observational biases.  Additionally, this new accretion estimate
agrees very well with the amount of matter that is needed to mildly
recycle a pulsar (Zdunik, Haensel \& Gourgoulhon 2002 on theoretical
calculations; Jacoby et al. 2005 on observational estimate).

This is not to say that the Brown scenario requiring twin binaries 
does not contribute to the double neutron star population.  But in 
our scenario with the reduced accretion rate, it is simply not the 
dominant formation scenario.  Because of this, our results are much 
less sensitive to the size of the twin population.

\section{Discussion}

Our calculations discussed in this paper involved a very simple twin
scenario, i.e. half of all the binaries were postulated to be equal
mass ($q>0.95$). In reality, the true fraction of twins is likely to
depend on the mass of the primary, most certainly on the orbital
separation of the two stars, and also possibly on the metallicity of
stars. Indeed, even the actual binary fraction is likely a function of
primary mass (e.g., Lada 2006). There was a recent report that the
fraction of B-type binaries in the LMC might be significantly lower
than in our Galaxy (Mazeh, Tamuz \& North 2006), indicating a
possibility of strong metallicity dependence in the efficiency of
binary formation.

From the absolute rates point of view, we show that effect of
including twins is relatively minor: although the merger rate does
indeed increase when twins are considered, the rate increase is fairly
small ($\sim1.5$). This is the direct result of evolutionary calculations 
that provide numerous channels of NS-NS formation without a strong
preference for the high mass ratio of progenitor binaries. The same 
calculations recover the empirically estimated rates of double neutron 
star mergers as some of their observed properties (Belczynski et al. 2007d).  
Also, chirp mass distribution for double compact objects formed with or 
without twins are almost indistinguishable. If double compact object are 
short-hard GRB progenitors, including twins in population synthesis 
calculations does not alter significantly the earlier rate predictions 
for the event rate.

Nevertheless, there are some interesting changes when we include
significant twin population. For one channel of binary evolution that
allows ultracompact binaries, introducing twins doubles the rate of 
very prompt NS-NS mergers (time to merger less than $10^{6}$ years)
compared to models with the ``flat'' $q$ distribution. In that
specific case, $\sim 70\%$ of all NS-NS binaries would merge within
$10^8\;$years after their formation (see Table~1), indicating a
possibility of a very significant population of short-hard gamma-ray
bursts associated with star forming galaxies. We should mention that
twins are not necessary to have a significant prompt population of
NS-NS binaries.  Even using most conservative assumptions,
$\sim 35\%$ of all NS-NS binaries merge within $10^8\;$years after
their formation (See Table~1). This is very interesting because of the
well localized short-hard bursts, roughly 40\% occured in young hosts,
15\% in old hosts, and location of roughly 45\% is still unknown
(Berger et al. 2007; E. Berger, private communication). So not only
are the ``prompt'' short-hard GRBs very common, they might turn out to
be the majority of this class of bursts, as the rough current limits
are from 40\% to even 85\%.

The fact that so many short-hard burst are found in star-forming
galaxies may have far reaching consequences for our understanding of
binary evolution (if indeed short-hard GRBs are connected to double
compact object mergers). If further observations find that even higher
fraction of short-hard GRBs is in young galaxies, that will indicate
strongly that the ultracompact channel of the binary evolution does
indeed lead to double compact object formation, something that
otherwise is very hard to resolve observationally. If that fraction is
higher still, i.e. 70\% and above, we will not only need the
ultracompact channel to be allowed, we might also need a significant
binary twin population to explain such a high rate (see
Table~1). Needless to say, more observational constraints on short-hard 
GRBs and their hosts are needed here.

\acknowledgements
We express special thanks to Marc Pinsonneault and Edo Berger for many 
useful comments. KB thanks members of the OSU Astronomy Department for 
their hospitality and numerous discussions.

\clearpage

\begin{deluxetable}{lccccl}
\tablewidth{360pt}
\tablecaption{Double Compact Objects: Flat (Twin) Populations\tablenotemark{a}}
\tablehead{Model\tablenotemark{b} & $R_{\rm Gal}$ [Myr$^{-1}$] & $R_{\rm A,ligo}$ 
           [yr$^{-1}$] & $F_{\rm 100}$ [\%] & $F_{\rm 1000}$ [\%] & Comments }
\startdata
A & 13 (20) &  15   (22) & 38 (36) & 72 (72) & reference model \\
B & 48 (74) & 692  (825) & 60 (71) & 83 (88) & HG CE allowed \\
\enddata
\label{rates}
\tablenotetext{a}{Values are given for calculations that account either for
flat or twin (in parenthesis) initial mass ratio distribution.} 
\tablenotetext{b}{We list Galactic merger rates ($R_{\rm Gal}$) and advanced 
LIGO detection rates ($R_{\rm A,ligo}$) for all double compact objects,
while fractions of mergers ($F_{\rm 100}$, $F_{\rm 1000}$) that take place in 
young galaxies ($<100$, $<1000$ Myr, respectively) are given only for potential 
GRB progenitors (i.e., NS-NS and BH-NS mergers).} 
\end{deluxetable}

\begin{figure}
\includegraphics[width=1.1\columnwidth]{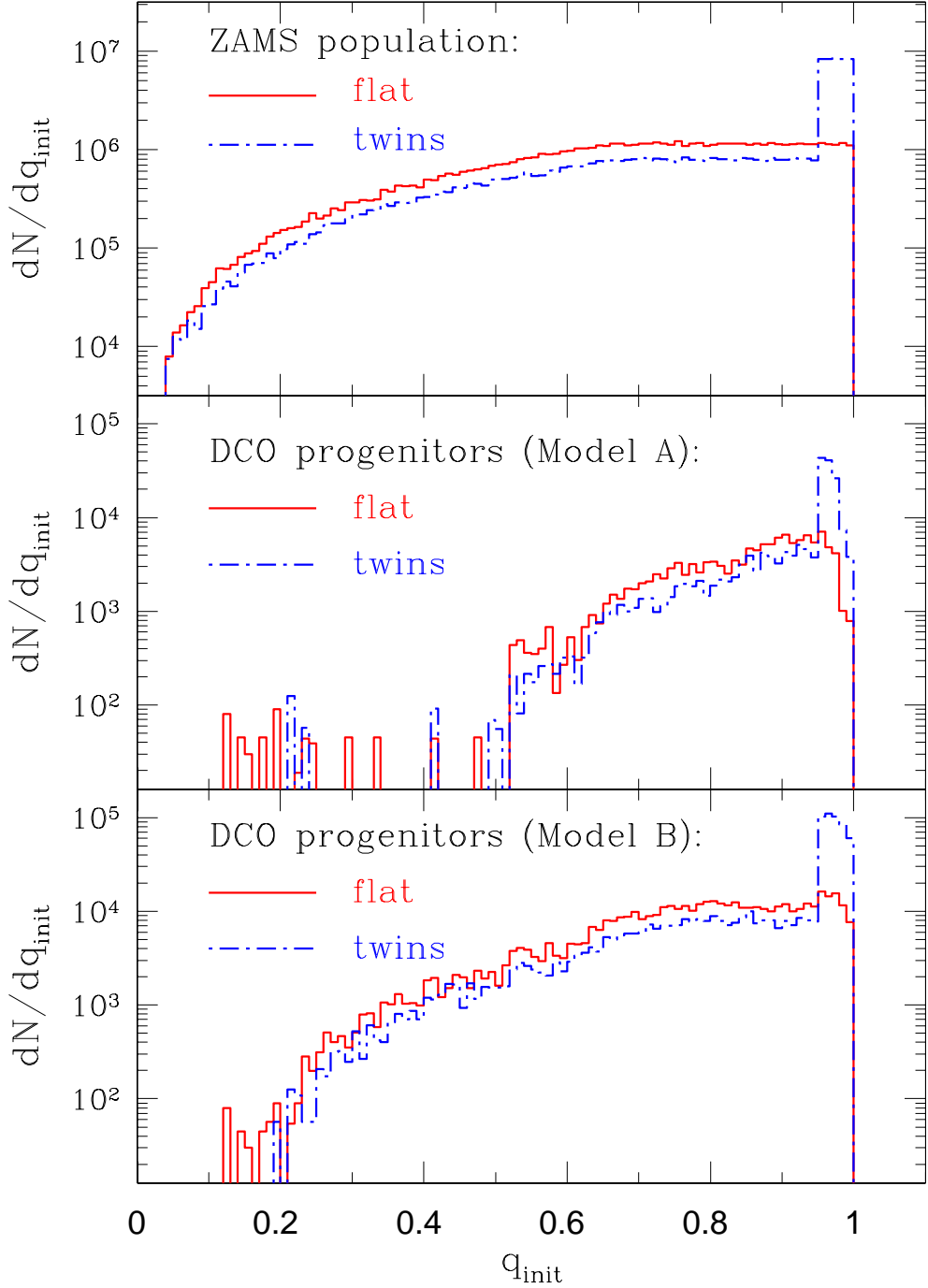}
\caption{Initial mass ratio distribution for Zero Age Main Sequence
binaries that we evolve (ZAMS; top panel) and the subpopulation of the 
above that in the end forms coalescing double compact objects (DCO; bottom 
panel). We show results of our two calculations: one with an underlying flat 
mass ratio distribution and the twin one with mass distribution peaked at 
high $q_{\rm init}$-values (for details see \S\,2.3). Note the change of 
vertical scale from the top panel to bottom panels. The numbers correspond 
to the entire Galactic population of double compact objects and their 
progenitors. 
}
\label{qdis}
\end{figure}
\clearpage

\begin{figure}
\includegraphics[width=1.1\columnwidth]{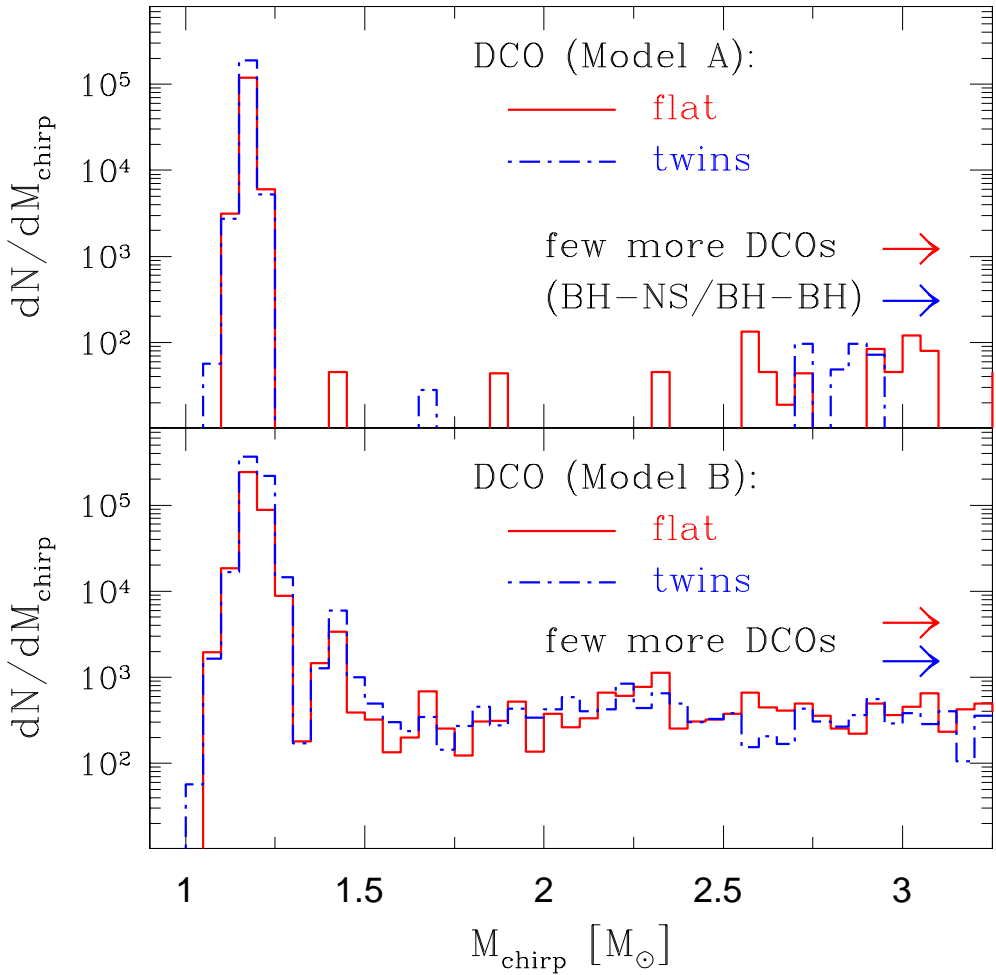}
\caption{Chirp mass distribution for double compact objects for model A
(top panel) and model B (bottom panel). 
Note that the flat and twin population distributions for double neutron stars 
(majority of DCOs) are almost the same and they peak at $\sim 1.2 \msun$, 
while only a small fraction of heavy compact objects (BH-NS and BH-BH binaries) 
extends to the chirp mass as high as $\sim 10 \msun$ (not shown).  
}
\label{Mtot}
\end{figure}
\clearpage

\begin{figure}
\includegraphics[width=1.1\columnwidth]{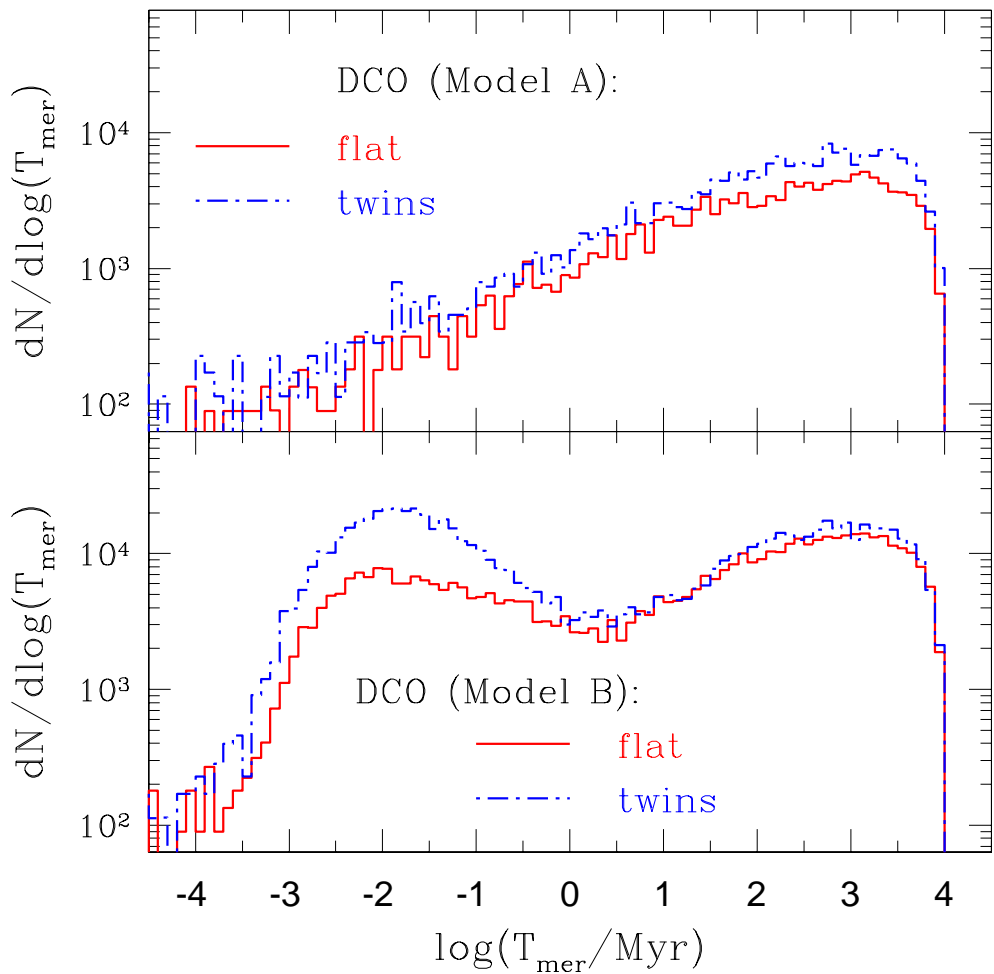}
\caption{Distribution of merger times for double compact objects for model 
A (top panel) and model B (bottom panel).
We can see that twin and flat populations are almost the same for model A
and that significant fraction of binaries ($\sim 35\%$) have merger times 
shorter than 100 Myr.
Model B with an alternative approach to common envelope evolution in which 
we allow survival of systems with Hertzsprung gap donors leads to formation 
of ultracompact double neutron stars that contribute to the short merger 
time peak ($T_{\rm mer}< 1$ Myr). 
We note that in this model even more: $\sim 60\%$ and $\sim 70\%$ DCOs form 
with short merger times for flat and twin mass ratio distributions, 
respectively. 
The short-lived systems are natural candidates for prompt short-hard GRBs
observed in young host galaxies.
}
\label{Tmr}
\end{figure}
\clearpage

\begin{figure}
\includegraphics[width=1.1\columnwidth]{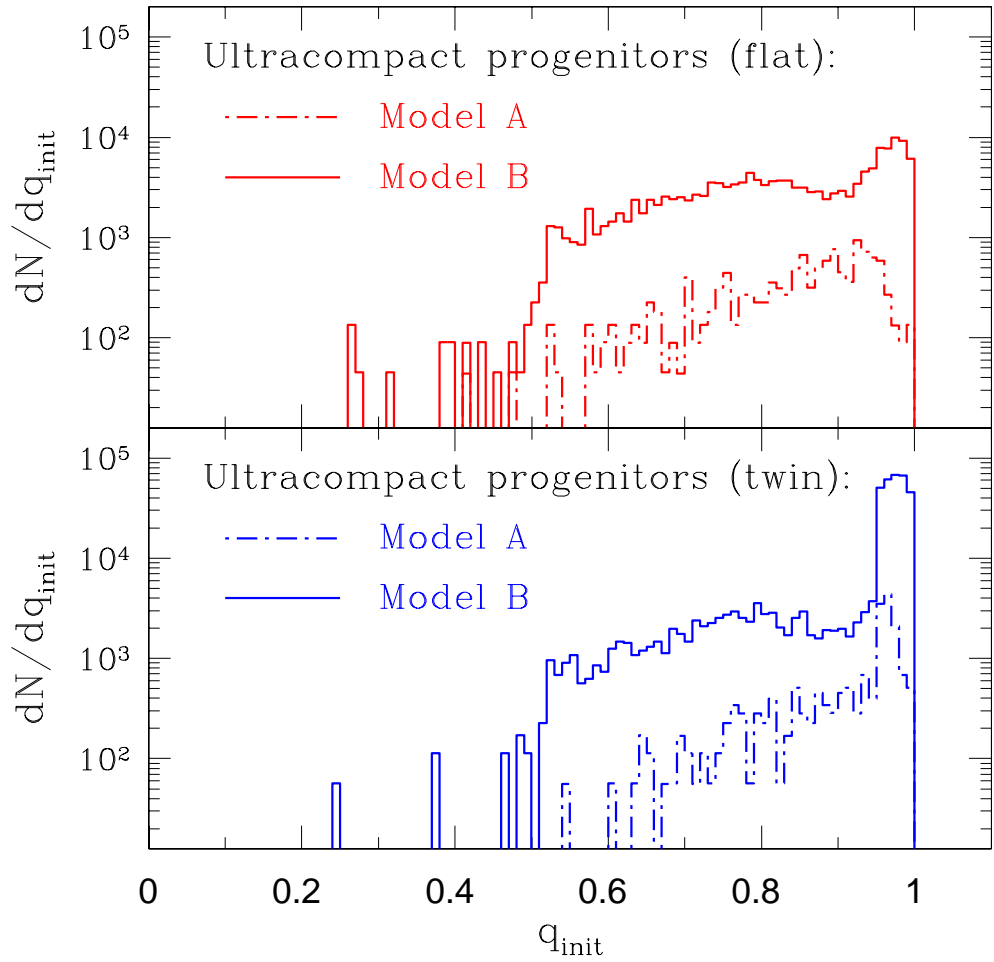}
\caption{Initial mass ratio distribution for Zero Age Main Sequence
binaries that evolve into ultracompact DCOs (merger times shorter than 
1 Myr). Results are shown for model A and B for both flat (top panel) 
and twin (bottom panel) populations. For more details see \S\,3.3.  
}
\label{qultra}
\end{figure}
\clearpage

\end{document}